\documentclass[%
 reprint,
nofootinbib,
 amsmath,amssymb,
 aps,
 pra,
]{revtex4-2}

\usepackage{graphicx}% Include figure files
\usepackage{dcolumn}% Align table columns on decimal point
\usepackage{bm}% bold math
\usepackage{hyperref}% add hypertext capabilities
\usepackage{mathtools}

\usepackage[colorinlistoftodos,prependcaption,textsize=footnotesize]{todonotes}

\DeclarePairedDelimiter\bra{\langle}{\rvert}
\DeclarePairedDelimiter\ket{\lvert}{\rangle}

\begin{document}

\title{Multipartite entanglement in the quantum tetrahedron}

\author{Robert Amelung}
\email{robert.amelung@fau.de}

\author{Hanno Sahlmann}%
\email{hanno.sahlmann@fau.de}
\affiliation{%
Institute for Quantum Gravity (IQG), Department of Physics,\\
Friedrich-Alexander-Universität Erlangen-Nürnberg (FAU),\\
Staudtstr. 7, 91058 Erlangen, Germany
}%

\date{\today}

\begin{abstract}
The space Inv$(j_1,j_2,j_3,j_4)$ of SU(2)-invariant four-valent tensors, also known as intertwiners, can be understood as the quantum states of a tetrahedron in Euclidean space with fixed areas. In loop quantum gravity, they are states of the smallest ``atom of space" with non-zero volume. At the same time they correspond to four-party tensor product states invariant under global rotations. 
We consider the multipartite entanglement of states in Inv$(j_1,j_2,j_3,j_4)$ using the recently proposed entropic fill. 

Numerically evaluating entropic fill in the case of equal spins between $1/2$ and $11$, we find that the distributions of entanglement are very different for intertwiners as compared to generic tensors, and for coherent intertwiners as compared to generic ones.
The peak in the distribution seems to be at the highest entanglement for generic intertwiners and at the lowest for generic tensors, but in terms of average entanglement, the roles are switched: average entanglement is highest in arbitrary tensors and lower in intertwiners, at least in the regime of large $j$. 

We also consider the dependence of entanglement on the geometric data of coherent intertwiners. Coherent intertwiners labeled by a symmetric tetrahedron have maximal entanglement, while minimal entanglement implies a coherent state labeled by a degenerate tetrahedron. Away from these cases, entanglement depends on the geometry in an intricate way. 

\end{abstract}

\maketitle

\section{Introduction}
Relations between entanglement and space-time geometry have been observed, investigated and even postulated in many contexts. A prime example is black hole entropy, which is associated with the geometric quantity (area) but may be understood as entanglement entropy between the outside and the inside of the black hole \cite{Page:1993wv,Bombelli:1986rw,Perez:2014ura,Srednicki:1993im}. More generally, relations between entanglement and geometry have been investigated in quantum field theory, see for example \cite{Srednicki:1993im}. In quantum gravity, entanglement is conjectured to play an important role \cite{Bianchi:2012ev}. In loop quantum gravity, various aspects of the interplay of geometry and entanglement have been investigated, see for example \cite{Livine:2017fgq,Baytas:2018wjd,Bianchi:2018fmq,Gruber:2018lef,Bianchi:2023avf,Sahlmann:2025ffi}. It has even been proposed that geometry itself is a form of entanglement \cite{Maldacena:2013xja}.

In the present work, we are considering a very simple geometric system, the quantum tetrahedron \cite{Barbieri:1997ks,Baez:1999tk,Freidel:2010xna}, and study the entanglement of its states. The smallest ``atom of space" in loop quantum gravity \cite{Thiemann:2001gmi,Ashtekar:2004eh,PolyhedraInLQG,Haggard:2023tnj} can be understood as a quantum tetrahedron, but the system can also be defined and studied independently. Its state space is a subspace of the space of four qudits,  
\begin{equation}
   \text{Inv}(j_1,j_2,j_3,j_4) \subset  \mathbb{C}^{d_{j_1}}\otimes \mathbb{C}^{d_{j_2}}\otimes\mathbb{C}^{d_{j_3}}\otimes\mathbb{C}^{d_{j_4}}, 
\end{equation}
where $d_j = 2j+1$ and Inv denotes the invariants under the action of SU(2) in the $j_1\otimes \ldots \otimes j_4$ representation. This invariance can be understood as a closure condition for the faces of the quantum tetrahedron \cite{Freidel:2010xna,PolyhedraInLQG}. 

The geometric properties of the tetrahedron correspond to operators on this space \cite{Ashtekar:1996eg,Ashtekar:1997fb,Major:1999mc,PolyhedraInLQG}. Since various geometric operators do not commute with each other, the geometries have fluctuations and generically behave quantum mechanically. There are, however, special semiclassical states that are thought to describe the geometry of a classical tetrahedron geometry in an optimal way: the coherent intertwiners \cite{Freidel:2010xna,NewSpinFoamVertexForQuantumGravity,Alesci:2016dqx}.  

Since this is a multiparty system, entanglement can be considered in various ways. Of particular interest is to quantify the genuine multipartite entanglement (GME), i.e., entanglement that goes beyond the bipartite entanglement between pairs of spins. In the present work we will use an intriguing measure of GME for a four-party system: the entropic fill \cite{EntropicFill}.
This GME measure is of particular interest as it unifies bipartite entanglement entropies in a geometric way. It generalizes to four parties the earlier \textit{concurrence fill} defined for three-party systems~\cite{TriangleMeasureofTripartiteEntanglement}. Entropic fill is a function of the volume of a tetrahedron, suitably normalized, the geometry of which is fixed by entanglement entropy of various subsystems. The areas of the four faces are given by the values of the entropies obtained when separating the system into one and three parties (\textit{one-to-other} entropies). Such a tetrahedron always exists thanks to proven entanglement polygon inequalities~\cite{EntanglementPolygonInequalityInQubitSystems,EntPolyQudits}: the sum of any three one-to-other entropies is greater than or equal to the remaining one. Entropies for the three possible partitions into two parties of two (\textit{two-to-two} entropies) are used to fix the two remaining degrees of freedom of such a tetrahedron. An issue of unique solvability (discussed below) is, to our knowledge, not conclusively resolved, but all numerical evidence suggests that entropic fill is well-defined for four-qubit and also four-qudit systems. Given that the quantum system under study is the quantum tetrahedron, a natural question to ask is how its geometry is related to that of the ``entropic tetrahedron'' used to calculate the entropic fill. It will be seen in the following that the answer is complicated but interesting. 

Entanglement of states of the quantum tetrahedron has been studied before. First results in this direction, on the many-to-many entanglement in mixtures of intertwiners, were obtained in \cite{Livine:2005mw,Livine:2005tfa}. Later, 
for example, the question was posed (and answered in the negative) whether there are intertwiners of valence four and higher that are absolutely maximally entangled \cite{Li:2016eyr,Mansuroglu:2022xey,Otto:2025fpc}. While the present work was underway, the interesting publication~\cite{DasCoherentIntertwinerEntanglementPaper} appeared. The authors investigate four-valent intertwiners systematically regarding their two-to-two entanglement: It is studied analytically, determined numerically for the group averaging of various tensors and several coherent tensors. 

The present work extends the consideration of \cite{DasCoherentIntertwinerEntanglementPaper} in various ways. The aim is 
to consider and compare distributions of entropic fill in several ensembles of states, and in relation to the quantum geometry embodied in the states. To characterize and compare the entropic fill, we show and discuss the results of three different numerical approaches:
\begin{itemize}
    \item Uniform random sampling of the various ensembles of four-party states with local dimension \(2j+1\).
    \item Visualization of entropic fill on the full configuration space of coherent intertwiners for classical geometric data corresponding to a tetrahedron.
    \item Two particular investigations of coherent intertwiners for classical geometric data not satisfying the closure condition.  
\end{itemize}
Using these approaches, we study the distribution of entropic fill in various ensembles, and for various local dimensions. In particular, we determine average entropic fill in the various cases. We also investigate entropic fill as a function of the tetrahedral geometry for coherent intertwiners, and as a function of the closure defect\footnote{i.e., a measure of the failure of coherent intertwiner labels to satisfy the closure condition.} for coherent intertwiners with non-geometric labels.

In section \ref{se:background} we briefly introduce the quantum tetrahedron, coherent intertwiners and the entropic fill, and describe our numerical pipeline. Section \ref{se:results} contains the results of the investigation. 

In the following, we work in units in which
\begin{equation}
    8\pi \beta \hbar G \equiv 8\pi \beta \ell_{P}^2 = 1\,, 
\end{equation}
where $G$ is Newton's constant, $\beta$ is the Barbero-Immirzi parameter of loop quantum gravity \cite{Thiemann:2001gmi,Ashtekar:2004eh}, and $\ell_{P}$ is the Planck length. 

\section{Four-valent intertwiners and entropic fill}
\label{se:background}

\textbf{Intertwiners} -- 
The results of the present work concern entanglement in intertwiner states, also known as invariant tensors, or equivariant maps. They are important in the recoupling theory of quantum mechanical spin. In loop quantum gravity, they are an important ingredient in constructing gauge-invariant states.  
Intertwiners can be understood as a linear subspace of multipartite spin states with a special ``invariance" property. This property demands that the state be unchanged under simultaneous \(\mathrm{SU}(2)\) rotations of all spins. Letting \(j_1,\dots,j_N\) be the spin quantum numbers for the \(N\) parties, the invariance of intertwiner \(I\) can be expressed as
\begin{equation}
\label{eq:invariance}
		(D^{(j_1)}(g)\otimes\dots\otimes D^{(j_N)}(g))\ I = I
\end{equation} 
for all \(g\in\mathrm{SU}(2)\), where \(D^{(j)}\) is the \(d_j\)\nobreakdash-dimensional irreducible representation (irrep) of \(\mathrm{SU}(2)\). In terms of the generators $\vec{J}_i$ on the $i$th tensor factor, in the irrep $j_i$, Eq.~\eqref{eq:invariance} is equivalent to 
\begin{equation}
    \label{eq:invariance_generators}
    \left(\sum_i \vec{J}_i\right)\, I = 0\,. 
\end{equation}
Therefore, such states are also equivalently described as having a total vanishing spin. Thus, they only exist if the \(N\) spins can combine to a total spin zero according to quantum-mechanical addition of angular momenta.

For two and three parties, the invariant subspace is at most one-dimensional. Starting from four parties, however, it has  several dimensions in general. In the case of equal spins \(j_1=j_2=j_3=j_4=:j\), the invariant subspace is \(d_j\)\nobreakdash-dimensional, just as every individual spin subsystem.\\

In loop quantum gravity, intertwiners can be thought of as quantum states of `grains' or `atoms' of quantized spatial geometry. While classical geometric intuition will generally fail, one possibility is to view $n$-valent intertwiners as quantum states of polyhedra embedded in flat space. In particular, the four-valent intertwiners can be thought of as states of a tetrahedron. To be more precise, denote by $A_i$ the area of the $i$th face of a polyhedron, and by $\vec{n}_i$ its unit normal, consideration of geometric operators in quantum gravity suggests to relate  
\begin{equation}
    \sqrt{\vec{J}_i^{\,2}} \,  \longleftrightarrow \,A_i\,, \qquad   \frac{\vec{J}_i}{\sqrt{\vec{J}_i^{\,2}} } \, \longleftrightarrow \, \vec{n}_i\,.
\end{equation}
The invariance conditions \eqref{eq:invariance} or \eqref{eq:invariance_generators} therefore correspond to the closure of the polyhedron via one of the Minkowski theorems on polyhedra \cite{PolyhedraInLQG}
\begin{equation}
    \sum_i A_i \,\vec{n}_i = 0\,. 
\end{equation}
In this picture, the intertwiners of irreducible representations are eigenstates of areas of the polyhedron, $A_i = \sqrt{j_i(j_i+1)}$, but not of its entire shape. Angles between surface normals in particular do not all commute with each other and are therefore not simultaneously diagonalizable. This is one motivation for the introduction of intertwiner states that describe a classical polyhedron geometry in the sense of distributing the fluctuations evenly among the degrees of freedom, which we describe in the following.
\\

\textbf{Coherent intertwiners} -- Semiclassical states of quantum geometry are \textit{coherent intertwiners}. They can be understood as the projections of products of Perelomov coherent states \cite{Perelomov:1971bd} onto the invariant subspace~\cite{NewSpinFoamVertexForQuantumGravity,Freidel:2010xna}. For a unit vector
\begin{equation}
		\vec{n} = (\sin\theta\,\cos\varphi,\ \sin\theta\,\sin\varphi,\ \cos\theta)\,,
\end{equation}
the corresponding coherent spin state is an \(\mathrm{SU}(2)\)-rotated version
\begin{align}
\begin{split}
		\lvert{j,\vec{n}}\rangle &:= D^{(j)}(g(\vec{n}))\, \lvert{j,j}\rangle\,,\\
		g(\vec{n})&:= \exp(\theta\,\vec{m}\cdot \vec{\tau})\,,\\
		\vec{m}&:= (-\sin\varphi,\ \cos\varphi,\ 0)\,
\end{split}
\end{align}
of the highest-weight spin state \(\lvert j,j\rangle\), where \(\vec{\tau}\) are the standard anti-Hermitian generators of \(\mathrm{SU}(2)\). Its components in the non-rotated spin basis can be expressed as~\cite{NewSpinFoamVertexForQuantumGravity}
\begin{align}
\label{eq:compontentsCoherent}
\begin{split}
		\langle{j,m}\rvert{j,\vec{n}}\rangle &= \sqrt{\frac{(2j)!}{(j-m)!\,(j+m)!}}\ \frac{\xi^{j-m}}{(1+|\xi|^2)^j}\,,\\
		\xi&:= \exp(-i\varphi)\,\tan\frac{\theta}{2}\,.
\end{split}
\end{align}
A coherent intertwiner \(I_\text{C}\in\mathrm{Inv}(j_1,j_2,j_3,j_4)\) corresponding to four unit vectors \(\vec{n}_i\) is defined as the \(\mathrm{SU}(2)\)\nobreakdash-average
\begin{equation}
    I_\text{C}(\vec{n}_1, \vec{n}_2, \vec{n}_3, \vec{n}_4):= \int_{\mathrm{SU}(2)}\mathrm{d}g\ \bigotimes_{i=1}^4 D^{(j_i)}(g)\,\lvert{j_i,\vec{n}_i}\rangle
\end{equation}
of a product of four coherent spin states, which amounts to a projection to the invariant subspace. Such intertwiners are invariant (up to a phase factor) under a simultaneous \(\mathrm{SO}(3)\) rotation of all four vectors \(\vec{n}_i\). When they fulfill the \textit{closure condition}
\begin{equation}
    \label{eq:closure_coherent}
    j_1\vec{n}_1+j_2\vec{n}_2+j_3\vec{n}_3+j_4\vec{n}_4=\vec{0}\,,
\end{equation}
they can be understood as the outward unit surface normals of the intertwiner's tetrahedral picture. In keeping with the literature on coherent intertwiners, we have used $A_i \propto j_i$ in \eqref{eq:closure_coherent} instead of the standard formula for the area spectrum of a single puncture in LQG.\footnote{The two spectra differ by an $\hbar$ correction, they can be obtained from different ordering choices \cite{Alekseev:2000hf,Sahlmann:2015dna}, and coincide asymptotically.} 

We note that a fascinating reformulation of the quantum geometry underlying LQG in terms of spinors has been discovered, that also leads to new ways to work with coherent intertwiners \cite{Freidel:2010xna,Freidel:2010aq,Freidel:2010tt,Borja:2010rc}. While we will not make use of these methods in the present work, they may be very useful to continue the present investigation.

In the following, we restrict to the case of equal spins \(j_1=j_2=j_3=j_4=:j\), so that the local dimension is the same for all four parties, namely \(d_j=2j+1\). A four-valent intertwiner can thus be interpreted as a particular four-qu\textit{d}it state.
\\

\begin{figure}[h]
    \centering
    \includegraphics[width=0.9\columnwidth]{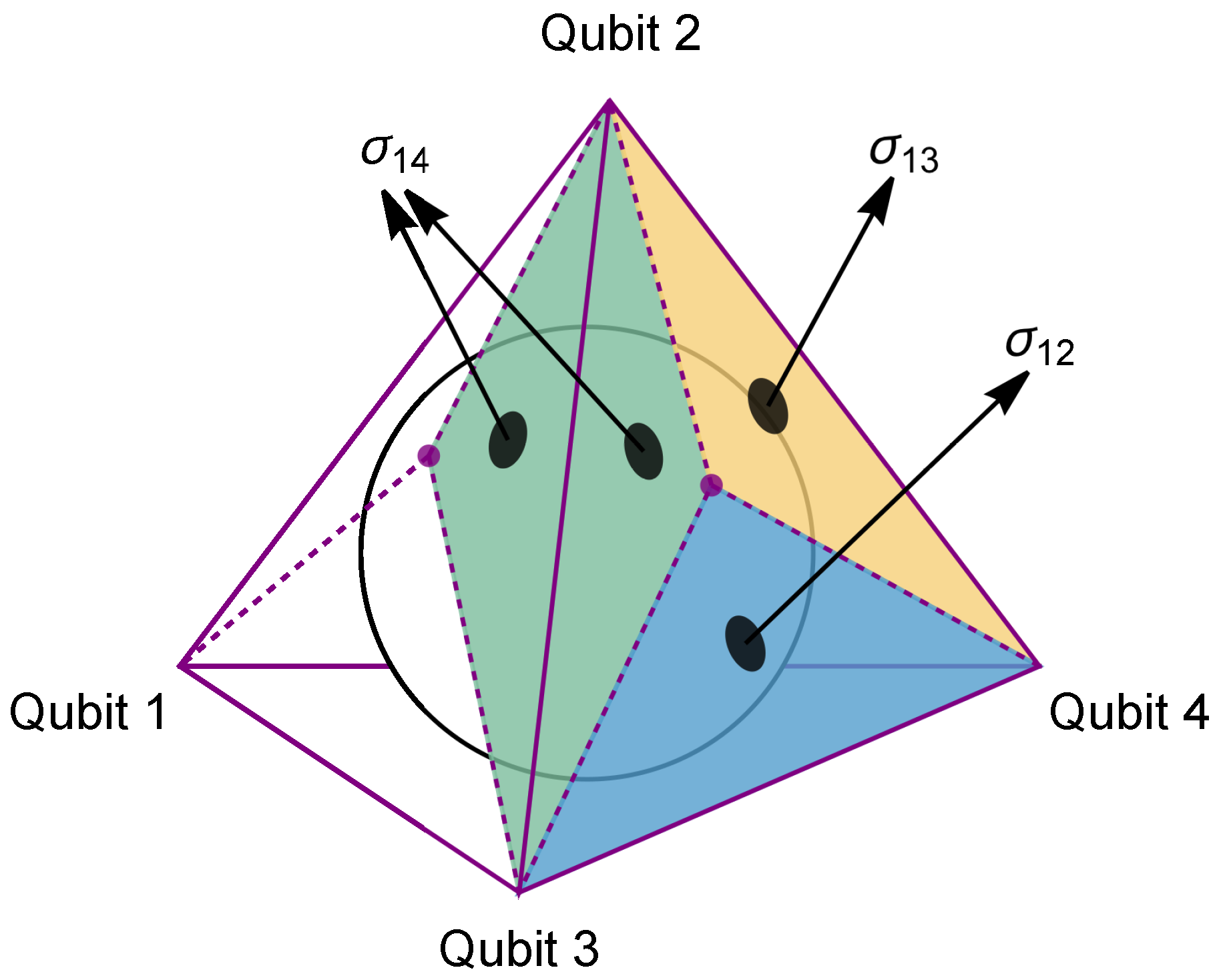}
    \caption{A tetrahedron with its inscribed sphere and the thus defined triangular areas on the faces. For each edge of the tetrahedron, the two adjacent triangles have the same area. The vertices are numbered so that $\sigma_{ij}$ is the area of each of the two triangles sharing the edge between the faces opposite vertices $i$ and $j$. This figure has been re-used without modification from~\cite{EntropicFill} under the license \href{https://creativecommons.org/licenses/by/4.0/}{CC BY 4.0}.}
    \label{fig:Tetrahedron_sigmas}
\end{figure}

\textbf{Entropic fill} -- The recently proposed \textit{entropic fill}~\cite{EntropicFill} permits quantitative comparisons of genuine multipartite entanglement in four-qubit systems. Generalizing the triangle \textit{concurrence fill} measure of tripartite entanglement~\cite{TriangleMeasureofTripartiteEntanglement}, an \textit{entropic tetrahedron} for four-qubit states is introduced in~\cite{EntropicFill}. 
Its geometry is determined by the bipartite entanglement entropies of the state, and its volume is the multipartite entanglement measure.   
The construction of the entropic tetrahedron relies on the property that every tetrahedron admits an inscribed sphere tangential to all four faces. By connecting the point of contact on each face to the three vertices of that face with straight lines, one segments the faces into three triangular regions each (for a total of 12; see Fig.~\ref{fig:Tetrahedron_sigmas}). For every edge of the tetrahedron, the two adjacent triangles have equal area. To index these triangles, one can first index the tetrahedron's vertices from 1 to 4. Following the convention in~\cite{EntropicFill}, we call $\sigma_{ij}$ the area of each of the two triangles sharing the edge between the faces opposite vertices $i$ and $j$. The six-parameter set \(\{\sigma_{12},\sigma_{13},\sigma_{14},\sigma_{23},\sigma_{24}, \sigma_{34}\}\) fixes the volume \(V\) of the tetrahedron by
\begin{align}
\begin{split}
		\label{eq:VolumeTetrahedron}
		V &= \frac{\sqrt{2}}{3}\ \sqrt{S}\ (A_0A_1A_2A_3)^{1/4}\,,\\*
		S &:= 2\cdot(\sigma_{12}+\sigma_{13}+\sigma_{14}+\sigma_{23}+ \sigma_{24}+ \sigma_{34})\,,\\
		A_0&:= +\sqrt{\sigma_{12}\sigma_{34}} + \sqrt{\sigma_{13}\sigma_{24}} + \sqrt{\sigma_{14}\sigma_{23}}\,,\\
		A_1&:= -\sqrt{\sigma_{12}\sigma_{34}} + \sqrt{\sigma_{13}\sigma_{24}} + \sqrt{\sigma_{14}\sigma_{23}}\,,\\
		A_2&:= +\sqrt{\sigma_{12}\sigma_{34}} - \sqrt{\sigma_{13}\sigma_{24}} + \sqrt{\sigma_{14}\sigma_{23}}\,,\\
		A_3&:= +\sqrt{\sigma_{12}\sigma_{34}} + \sqrt{\sigma_{13}\sigma_{24}} - \sqrt{\sigma_{14}\sigma_{23}}\,.
\end{split}
\end{align}
To determine the entropic fill of a (pure) four-qubit state, one associates the state with an ``entropic" tetrahedron. The \(\sigma_{ij}\) parameters of this tetrahedron are obtained from the bipartite \textit{entanglement entropies} of the seven possible bipartitions of the system: four \textit{one-to-other} entropies \(E_{i(jkl)}\) and three \textit{two-to-two} entropies \(E_{(ij)(kl)}\), defined using the base-two logarithm. The entropies fix the~\(\sigma_{ij}\) by the following sets of equations:
\begin{align}
		\label{eq:F4FirstSet}
		\sigma_{ij} + \sigma_{ik} + \sigma_{il} &= E_{i(jkl)}\,,\\*
		\label{eq:F4SecondSet}
		-\sqrt{\sigma_{ij}\sigma_{kl}} + \sqrt{\sigma_{ik}\sigma_{jl}} + \sqrt{\sigma_{il}\sigma_{jk}} &= \lambda\ E_{(ij)(kl)}\,.
\end{align}
These are understood for all index permutations (a total of seven non-equivalent equations), and \(\lambda\) is an additional non-physical variable, introduced to match the number of variables to the number of equations. The first set~\eqref{eq:F4FirstSet} enforces that the four one-to-other entropies are the respective areas of the faces of the entropic tetrahedron. This is always possible because of qubit entanglement polygon inequalities~\cite{EntanglementPolygonInequalityInQubitSystems}. The two-to-two entropies, on the other hand, determine the tetrahedron's overall shape given these constraints for the faces. Roughly speaking, the spread of the two-to-two entropies can be viewed as a measure of the tetrahedron’s irregularity. Indeed, when all four one-to-other entropies are equal, then the entropic tetrahedron is regular if all two-to-two entropies are equal as well. The non-physical variable \(\lambda\) removes the dependence on the absolute magnitudes of the two-to-two entropies, so that only their ratios are of importance.
Given the volume~\(V\) of the entropic tetrahedron determined from the obtained \(\sigma_{ij}\), the entropic fill measure \(F_4\) is defined as
\begin{equation}
    F_4 = (3^{7/6}/2)\,V^{2/3}\,,
\end{equation}
where the pre-factor is introduced for normalization \({0\leq F_4\leq 1}\), while the exponent permits additivity of the resulting measure.

For the entropic fill to be well-defined for all four-qubit states, Eqs.~\eqref{eq:F4FirstSet} and~\eqref{eq:F4SecondSet} must be uniquely solvable in the domain \(\sigma_{ij}\geq 0\) for any combination of entropies that a four-qubit state can exhibit (at least up to configurations that yield the same volume). To our knowledge, this has not yet been established rigorously. Nevertheless, the authors of~\cite{EntropicFill} report that their numerical investigations consistently support the existence of a unique solution with non-negative values, indicating that an entropic tetrahedron can always be constructed for four-qubit states. We accepted this as sufficient to use entropic fill in our numerical approach.

Assuming that entropic fill is well-defined (or restricting to those states for which it is), it has been shown to be a measure of \textit{genuine} multipartite entanglement (GME)~\cite{EntropicFill}. Foremost, this means that it assigns non-zero values to states that are not biseparable under any bipartition, and only to those. Often, additivity and local monotonicity are also required properties for a GME measure. For the entropic fill, all those properties have been established in \cite{EntropicFill}. 

As noted in~\cite{EntropicFill}, entropic fill can be suitably generalized to four-\textit{qudit} systems (i.e.\ the same local dimension \(d\geq2\) for each subsystem), since polygon inequalities hold in that case as well~\cite{EntPolyQudits} (for any four-qudit state, there exists a tetrahedron for which the areas of its faces are given by the one-to-other entropies of the state), up to the same unique solvability issue of Eqs.~\eqref{eq:F4FirstSet} and~\eqref{eq:F4SecondSet} already discussed in the four-qubit case. Since the maximal value of bipartite entropies depends on the local dimensions, we chose to normalize all entropies --~so that \(0\leq E\leq 1\)~-- before plugging into the equations. Since~\(V\) is homogeneous in \(\sigma_{ij}\), this is equivalent to dividing by the maximal~\(F_4\) at the end. Indeed, the maximal volume is obtained when all~\(\sigma_{ij}\) are equal to \(1/3\) of the maximal one-to-other entropy (yielding a regular entropic tetrahedron). Note that this normalization has no impact on the solvability of the equations.

The normalization of the two-to-two entropies is irrelevant, as it can be absorbed into \(\lambda\). This means that four-qudit states need not be \textit{perfect}, in the sense of maximal entanglement for all possible bipartitions, to exhibit maximal entropic fill. It suffices that the one-to-other entropies are maximal, while the two-to-two entropies all have the same value. Note that if at least one of the one-to-other entropies is non-zero, not all two-to-two entropies can vanish simultaneously due to sub-additivity~\cite{Ruskai}.

In the four-qudit case also, the authors in~\cite{EntropicFill} report numerical evidence supporting the unique solvability of Eqs.~\eqref{eq:F4FirstSet} and~\eqref{eq:F4SecondSet} up to local dimension~\(d=50\), conjecturing that \(F_4\) is well-defined for all four-qudit states. Our numerical findings support this conjecture for \(d\leq 23\): When numerically solving Eqs.~\eqref{eq:F4FirstSet} and~\eqref{eq:F4SecondSet} in samples of considerable size, we consistently obtained residual norms below \(10^{-9}\) and found the distributions to be stable under choosing different initial guesses, suggesting that the numerical solutions indeed approximate real unique solutions.\\

\textbf{Methods} -- We computed entropic fill entirely numerically. Given a general four-party state -- typically encoded as a four-dimensional component array -- we computed the seven (normalized) entanglement entropies using singular value decompositions. For intertwiners, only the two-to-two entropies needed to be determined, while the one-to-other entropies were set to unity as they are always maximal (see below). Using these values, we solved Eqs.~\eqref{eq:F4FirstSet} and~\eqref{eq:F4SecondSet} numerically to determine the parameters \(\sigma_{ij}\) of the entropic tetrahedron. The components of coherent intertwiners -- given the unit normals \(\vec{n}_1,\dots,\vec{n}_4\) -- were determined via numerical projection (followed by normalization) to the invariant subspace of the corresponding coherent product state, the components of which were computed in accordance with Eq.~\eqref{eq:compontentsCoherent}. For all results below, the obtained numerical solutions satisfy Eqs.~\eqref{eq:F4FirstSet} and~\eqref{eq:F4SecondSet} with Euclidean residual norms below \(10^{-9}\) for almost every sample, when choosing a regular entropic tetrahedron as initial guess. We found the results to be stable under variation of the initial guess, yielding residual norms almost never larger than \(10^{-8}\) when choosing the initial guess randomly.

\section{Results}
\label{se:results}

\textbf{General properties} -- The one-to-other entropies of intertwiners are always maximal~\cite{Li:2016eyr}
\begin{equation}
\label{eq:1-3entropies}
    E_{i(jkl)} = 1\,,
\end{equation}
because the reduced density matrix is an intertwiner from the $j$ irrep to itself and hence proportional to the identity. The entropic tetrahedra must hence be \textit{equifacial} for those states; such tetrahedra are also called \textit{disphenoids}. The two-to-two entropies \(E_{(ij)(kl)}\) of four-valent intertwiners with equal spins, however, cannot all be maximal simultaneously, since there exist no perfect four-valent invariant tensors for equal local dimension~\cite{Li:2016eyr}. Nevertheless, as a special case, the coherent intertwiner whose normals~\(j\vec{n}_i\) correspond to a regular tetrahedron is maximally entangled according to entropic fill. Indeed, the proper rotational symmetries of a regular tetrahedron contain, in particular, rotations cyclically permuting three of the four normal vectors. Since exchanging normal vectors correspondingly (inversely) permutes components of the tensor, its three two-to-two entropies must be identical, leading to maximal entropic fill as discussed. Note that general disphenoids only exhibit \(180\)° rotational symmetries which translate into pairwise -- not cyclic~-- exchanges of indices, so the entropic fill of coherent intertwiners with closure is not maximal in general.\\

\textbf{Sampled distributions} -- To compare the entropic fill among different types of four-spin states, we numerically sampled \(5\times 10^5\) randomly chosen tensors for each of four different cases at small values of~\(j\): arbitrary; invariant; coherent with and without a fulfilled closure condition. Arbitrary tensors were sampled uniformly with respect to the Fubini-Study metric from the unit sphere in~\(\mathbb{C}^{d_j^4}\); the components were interpreted in the standard spin product basis. Similarly, invariant tensors were sampled uniformly from the unit sphere in~\(\mathbb{C}^{d_j}\), interpreted within a specific orthonormal basis of \(\mathrm{Inv}(j^{\otimes 4})\). For the coherent intertwiners without closure, we independently chose four unit vectors in \(\mathbb{R}^3\), sampled uniformly from the unit sphere, and computed the corresponding tensor.

\begin{figure}[h]
    \centering
    \includegraphics[width=\columnwidth]{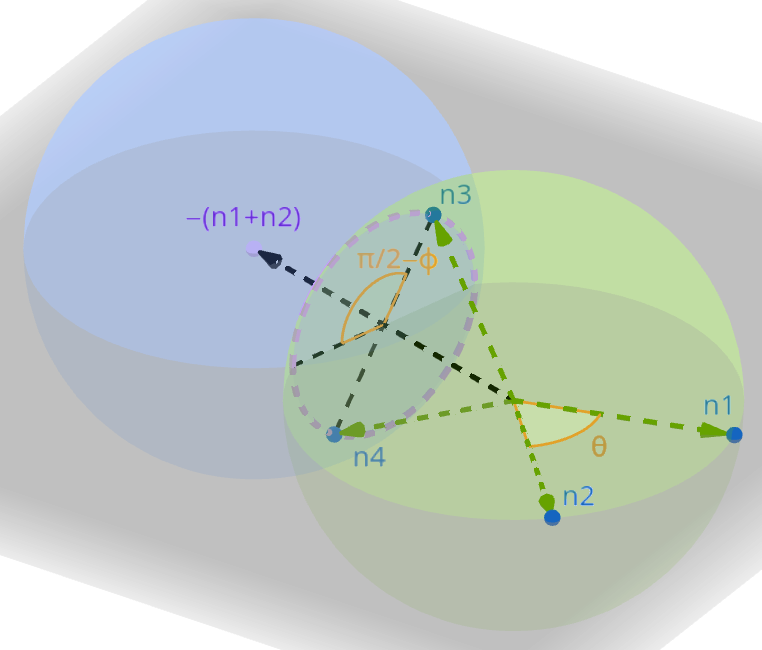}
    \caption{Example configuration of closing unit vectors \(\vec{n}_1,\dots,\vec{n}_4\) with associated angles \(\theta=\angle(\vec{n}_1,\vec{n}_2)\) and \(\phi\). Given \(\vec{n}_1\) and \(\vec{n}_2\), the other two unit vectors must point to antipodal points of the circle formed by the intersection between unit spheres centered at the origin and at \(-\vec{n}_1-\vec{n}_2\). The angle~\(\phi\) parametrizes the choice of \(\vec{n}_3,\vec{n}_4\) given the first two unit vectors; \(\pi/2-\phi\) is the oriented angle between the planes \((\vec{n}_1,\vec{n}_2)\) and \((\vec{n}_3,\vec{n}_4)\). Together,~\(\theta\) and~\(\phi\) encode the sets of four closing three-dimensional unit vectors, up to a general rotation.}
    \label{fig:theta_phi}
\end{figure}

However, to generate random unit vectors which close, one must proceed differently. The space of four-tuples of unit vectors -- modulo proper rotations~-- whose sum vanishes has two degrees of freedom; it is topologically a sphere and admits a natural uniform measure. It is a special case of the general Kapovich-Millson phase space~\cite{KapovichMillson1996}. As one of the two degrees of freedom, we chose the angle \(\theta\in(0,\pi)\) enclosed by \(\vec{n}_1\) and \(\vec{n}_2\). Given these two vectors, the endpoints of \(\vec{n}_3\) and \(\vec{n}_4\) are restricted to the intersection between the unit sphere centered at the origin and the unit sphere centered at \(-\vec{n}_1-\vec{n}_2\), which is a circle. We chose the angle \(\phi\in[0,2\pi)\) along this circle as the second parameter; \(\pi/2-\phi\) is the oriented angle between the planes \((\vec{n}_1,\vec{n}_2)\) and \((\vec{n}_3,\vec{n}_4)\). A visualization of these angles is shown in Fig.~\ref{fig:theta_phi}. The induced uniform probability distribution on the phase space dictates that~\(\theta\) be chosen according to the density \(\sin(\theta/2)/2\), while~\(\phi\) is selected uniformly~\cite{ClosedRandomWalks}. Note that in the non-closure case,~\(\theta\) follows a \(\sin(\theta)/2\) distribution. We sampled \(\vec{n}_1\) and \(\vec{n}_2\) accordingly, followed by determining \(\vec{n}_3\) on the circle and choosing \(\vec{n}_4=-\sum_{i=1}^3 \vec{n}_i\). Absolute orientation of the four-tuple is irrelevant for measures of entanglement.

Note that the sampling for the ensembles of coherent intertwiners is thus uniform in their parameter space, but does not make reference to the Fubini-Study volume.

Distributions for \(j\in\{0.5,1.5,3\}\) are shown in Fig.~\ref{fig:Distributions}, while means are shown as a function of \(j\) in Fig.~\ref{fig:Distributions_means}. The mean entanglement seems to increase fastest for arbitrary states and general intertwiners, out-performing coherent intertwiners.

Within the coherent category, the closure condition leads to slightly higher entropic fill for spin values up to about \(j=7\). Moreover, coherent intertwiners -- especially those with closure -- exhibit narrow peaks of comparably very likely entropic fill values. These features do not seem to be mere numerical artifacts. Indeed, the residual norms for the intertwiners responsible for the peaks are consistently very small (often on the order of \(10^{-11}\)), which suggests that the numerical solver used to calculate the entropic fill is unlikely to be artificially generating these peaks. Moreover, we found that the distributions of the two-to-two entropies of coherent intertwiners also exhibit pronounced peaks at the same entropy values attained by the intertwiners contributing to the entropic fill peak, suggesting that these features of the entropic fill are a consequence of the two-to-two entropies' distributions of coherent intertwiners when sampled in this way, and as such, a part of the entropic fill distribution. The structure of entropic fill as a function of the geometry of the coherent intertwiner's parametrizations (see next sub-section) is intricate but does not show any apparent irregularities or numerical artifacts. In fact, the value of the entropic fill at those features in Fig.~\ref{fig:Distributions} corresponds to a relatively large region in the coherent state parameter space. This gives these features further credibility. 

Arbitrary states exhibit a peak around their mean entropic fill, while general intertwiners have a clear increasing preference for near-maximal values. For coherent intertwiners, the distribution drops much slower to zero at lower entropic fill than the other two categories.\\

\textbf{Full configuration space} -- To gain understanding about the structure of entropic fill within the category of four-valent coherent intertwiners, one can analytically determine the two-to-two entropies as a function of the angles \(\theta_{ij}:=\angle(\vec{n}_i,\vec{n}_j)\). (The one-to-three entropies are always maximal, Eq.~\eqref{eq:1-3entropies}.) In~\cite{DasCoherentIntertwinerEntanglementPaper}, general formulas in terms of Clebsch-Gordan coefficients are derived, but for low~\(j\) one is able to find relatively simple expressions, at least for the relevant reduced density matrices. Indeed, \(\rho_{ij}:=\mathrm{tr}_{ij}\ket{I_\text{C}}\bra{I_\text{C}}\) is a four-valent intertwiner itself, and as such can be expanded in a basis of the \(d_j\)-dimensional invariant subspace, 
\begin{equation}
    \rho_{ij} = \sum_{k=0}^{2j} \lambda_k^{(i,j)} P_k^\dagger P_k\,, 
\end{equation}
where $P_k$ denotes the intertwiner to the $k$ irrep, written as a projection operator on $j\otimes j$. 
One sees that the components $\lambda_k^{(i,j)}$ are the eigenvalues of $\rho_{ij}$ and have multiplicity \(d_k\). If the norm of \(I_\text{C}\) is known, one only needs to compute \(2j\) eigenvalues.

For the case \(j=1/2\), we computed the eigenvalues as
\begin{align}
    \lambda_0^{(i,j)} &= \frac{1}{4}\sin^2\left(\frac{\theta_{ij}}{2}\right)\sin^2\left(\frac{\theta_{kl}}{2}\right)\,,\\
    \lambda_1^{(i,j)} &= \frac{\|I_\text{C}\|^2 - \lambda_0^{(i,j)}}{3}\,,
\end{align}
where \(k,l\) are the other two indices in the two-to-two bipartition \((ij)(kl)\). A general integral formula for the squared norm of \(I_\text{C}\) can be found in~\cite{NewSpinFoamVertexForQuantumGravity}, which for the case of four parties and \(j=1/2\) evaluates to
\begin{align}
\begin{split}
\|I_\text{C}\|^2
=
\frac{1}{6}
\Big[
\sin^2\frac{\theta_{12}}{2}\,\sin^2\frac{\theta_{34}}{2}
&+
\sin^2\frac{\theta_{13}}{2}\,\sin^2\frac{\theta_{24}}{2}\\
&+
\sin^2\frac{\theta_{14}}{2}\,\sin^2\frac{\theta_{23}}{2}
\Big]\,.
\end{split}
\end{align}
In summary, the normalized two-to-two entropies of four-valent coherent intertwiners with common spin \(1/2\) can be given as
\begin{alignat}{2}
\begin{split}
\label{eq:2-2entropiesspin0.5}
    E_{(ij)(kl)} &= -\frac{1}{2}\,\mathrm{tr}\Big[\frac{\rho_{ij}}{\|I_\text{C}\|^2}\log_2\Big(\frac{\rho_{ij}}{\|I_\text{C}\|^2}\Big)\Big] \\ 
    &= -\frac{1}{2\|I_\text{C}\|^2} \Big[ \lambda_0^{(i,j)} \log_2\Big(\frac{\lambda_0^{(i,j)}}{\|I_\text{C}\|^2}\Big)\\
    &\qquad\qquad\qquad+ 3\lambda_1^{(i,j)}\log_2\Big(\frac{\lambda_1^{(i,j)}}{\|I_\text{C}\|^2}\Big)\Big]\,,
\end{split}
\end{alignat}
which are explicit expressions only in terms of the angles \(\theta_{12},\theta_{13}, \theta_{14}, \theta_{23},\theta_{24}, \theta_{34}\). For the case of the coherent intertwiners \textit{with closure}, a slight simplification is possible. Indeed, opposite dihedral angles in disphenoids are equal, so we can set \(\theta_{12}=\theta_{34}\), \(\theta_{13}=\theta_{24}\), and \(\theta_{14}=\theta_{23}\) in this case. Furthermore, once two of these angles are known, the third can be inferred through the closure condition as
\begin{equation}
    \cos\theta_{14}=-1-\cos\theta_{12}-\cos\theta_{13}\,,
\end{equation}
which restricts any pair \((\alpha,\beta)\in[0,2\pi]^2\) of these angles by \(\left|\alpha-\pi\right|+\left|\beta-\pi\right|\le \pi\). Relating these angles to our chosen parametrization \((\theta,\phi)\) of the closing coherent intertwiner space is possible via
\begin{align}
    \cos\theta_{12} &= \cos\theta\,,\\
    \cos\theta_{13}
    &=-1+\sin^2\left(\frac{\theta}{2}\right)(1-\sin\phi)\,,\\
    \cos\theta_{14}
    &=-1+\sin^2\left(\frac{\theta}{2}\right)(1+\sin\phi)\,.
\end{align}
The task of analytically determining entropic fill remains difficult even for this special case, since Eqs.~\eqref{eq:F4FirstSet} and \eqref{eq:F4SecondSet} do not admit a straightforward general solution given the two-to-two entropies~\eqref{eq:2-2entropiesspin0.5}. We observe a special solution for \(j=1/2\), however: the entropy \(E_{(ij)(kl)}\) vanishes if and only if the trace of \(\rho_{ij}\) is fully concentrated in \(\lambda_0^{(i,j)}\), i.e., if \(\lambda_0^{(i,j)}=\|I_\text{C}\|^2\). For coherent intertwiners with closure, our expressions show that this is only the case for
\begin{equation}
\label{eq:F4is0}
    \theta_{ij}=\pi\,,\qquad \theta_{ik}\in\left\{\frac{\pi}{2}, \frac{3\pi}{2}\right\}
\end{equation}
within the allowed region, which represent a degenerate tetrahedron configuration. Its geometric volume $V_3$ is vanishing. Since entropic fill is a GME measure, \(F_4\) vanishes only at these degenerate points. We have 
\begin{equation}
 F_4 = 0 \quad \Rightarrow \quad V_3 =0 \qquad (\text{for } j=1/2).  
\end{equation}
The converse does not hold, however, as  
there are other configurations (e.g.\ $\theta_{12}=\theta_{34}=0,\ \theta_{13}=\theta_{24}=\pi$) for which $V_3=0$ but $F_4 \neq0$.    

In Fig.~\ref{fig:CohClosure_FullConfigSpace_TH12_TH13_j0.5}, we show the entropies from these formulas, as well as the numerically obtained entropic fill, as functions of~\(\theta_{12}\) and~\(\theta_{13}\). As expected, we observe distinct valleys for each of the three two-to-two entropies, which directly translate to dips in entropic fill. At the bottom of these valleys, \(F_4\) vanishes.

The above analytical calculations can, in principle, be repeated analogously for arbitrary spin \(j\), but we expect the expressions to only grow in complexity. Since we confirmed that these analytical results for the entropies coincide with our fully numerical approach, we restrict in the following to purely numerical results.

We show numerically computed entropies and entropic fill in the full \((\theta,\phi)\) configuration space of coherent intertwiners with closure, in Fig.~\ref{fig:CohClosure_FullConfigSpace_j0.5} for \(j=1/2\) and Fig.~\ref{fig:CohClosure_FullConfigSpace_j1.5} for \(j=3/2\). We evaluated \(300\times 300\) regularly spaced points in the configuration space. For this, \(\vec{n}_1\) was set to the unit vector in the \(y\)-direction, and \(\vec{n}_2=(\sin\theta, \cos\theta,0)\). We chose coordinate unit vectors for the plane of the circle of possible \(\vec{n}_{3,4}\)\nobreakdash-endpoints, namely the \(\hat{z}\) unit vector together with \({(\vec{n}_1+\vec{n}_2)/||{\vec{n}_1+\vec{n}_2}||\times \hat{z}}\). Then the angular parameter \(\phi\in[0,2\pi)\) for \(\vec{n}_3\) corresponds to turning around the circle in the positive sense. Note that the endpoints of \(\vec{n}_3\) and \(\vec{n}_4\) are always antipodal on this circle, so \(E_{13}\) and \(E_{14}\) are equal up to a translation \(\phi\to\phi+\pi\). Note also that area elements in these plots are not equally likely for a random coherent intertwiner with closure; in fact,~\(\theta\) near~\(\pi\) is more likely than near zero.

Overall, we observe dips in single two-to-two entropies (blue regions) that result in corresponding low-entropic-fill valleys, surrounded by ``stripes" of high entropic fill. For small values of \(\theta\), \(\vec{n}_3\) and \(\vec{n}_4\) are restricted to a small circle, explaining the entropies' and entropic fill's approximate independence of \(\phi\) in that region. This independence even goes up to values above \(\theta=\pi/2\) for~\(E_{12}\). Furthermore, the regular tetrahedron configurations (\(\theta=\arccos(-1/3)\), \(\phi\in\{0,\pi\}\)) show as islands of maximal entropic fill. The peaks in Fig.~\ref{fig:Distributions} correspond to (parts of) the white stripes. 

We also show the mean entropies and entropic fill of coherent intertwiners with closure as a function of~\(\theta\) in Fig.~\ref{fig:Means_given_theta} for \(j\in\{0.5,1.5,3\}\). For this, we sampled \(500\) equally spaced values of~\(\phi\) for each of \(300\) equally spaced values of~\(\theta\). Note that since \(E_{13}\) and \(E_{14}\) are related by \(\phi\)-translational symmetry, their means coincide. We observe dips in mean entropic fill around \(\theta=\pi/2\) and near \(\theta=\pi\), whose depth decreases as \(j\) increases. We observe no particular structure near the regular tetrahedral angle \(\theta\approx1.91\). The mean two-to-two entropy \(\overline{E_{12}}\) exhibits a peak near \(\theta=\pi/2\), which agrees with numerical results in~\cite{DasCoherentIntertwinerEntanglementPaper}.\\

\textbf{Particular configurations} -- As a final investigation into the entanglement of coherent intertwiners, we chose two particular \textit{base configurations} -- a regular tetrahedron, and a \textit{tetragonal} disphenoid (isosceles faces) -- and numerically computed the entropic fill as a function of the first unit vector, while the other three remained fixed. Closure is thus obtained only at one particular \(\vec{n}_1\). The following base configuration polar coordinates were chosen for the regular tetrahedron:
\begin{align*}
\begin{split}
    (\varphi_1,\varphi_2,\varphi_3,\varphi_4)&=(-\frac{\pi}{4}, -\frac{3\pi}{4}, \frac{\pi}{4}, \frac{3\pi}{4})\,,\\
    (\cos\theta_1,\cos\theta_2,\cos\theta_3,\cos\theta_4)&=(-\frac{1}{\sqrt{3}}, \frac{1}{\sqrt{3}}, \frac{1}{\sqrt{3}}, -\frac{1}{\sqrt{3}})\,.
\end{split}
\end{align*}
For the tetragonal disphenoid, we set
\begin{align*}
\begin{split}
    (\varphi_1,\varphi_2,\varphi_3,\varphi_4)&=(-\frac{\pi}{2}, \frac{3 \pi}{4}, -\frac{\pi}{2}, \frac{\pi}{4})\,,\\
    (\theta_1,\theta_2,\theta_3,\theta_4)&=(\frac{\pi}{4}, \frac{\pi}{2}, \frac{3\pi}{4}, \frac{\pi}{2})\,.
\end{split}
\end{align*}
The results are shown in Fig.~\ref{fig:RegTet_Disph_j0.5} for \(j=1/2\), and in Fig.~\ref{fig:RegTet_Disph_j1.5} for \(j=3/2\), where we sampled a regular grid of \(300\times 300\) pairs of \(\cos\theta_1\) and \(\varphi_1\). In these plots, every area element is equally likely when choosing \(\vec{n}_1\) uniformly on the unit sphere.

In absolute terms (left plots), we see clear relative dips in entropic fill at three distinct coordinate pairs. When plotting logarithmically (right plots), we see two peaks of near-maximal entropic fill. For the regular tetrahedron, one of them lies at the base configuration as expected, but there is a second one precisely at its antipodal point. This likely originates from the high degree of symmetry in this configuration, and for \(j=1/2\) our analytical expressions confirm this observation.
We additionally note that the entropic fill becomes small (and $=0$ in the case of $j=1/2$) when the orientation of two unit normals is flipped relative to the closed regular tetrahedron. 

For the tetragonal disphenoid, the peaks are very narrow. These examples suggest interesting entanglement structure within near-closing coherent intertwiners.

\section{Discussion and outlook}
In the present work we have used entropic fill to characterize genuine multipartite entanglement in intertwiners, i.e., tensors invariant under rotations. This continues and complements other works \cite{Li:2016eyr,Mansuroglu:2022xey,Otto:2025fpc,DasCoherentIntertwinerEntanglementPaper}. 

\begin{figure*}
    \centering
    \includegraphics[width=\textwidth]{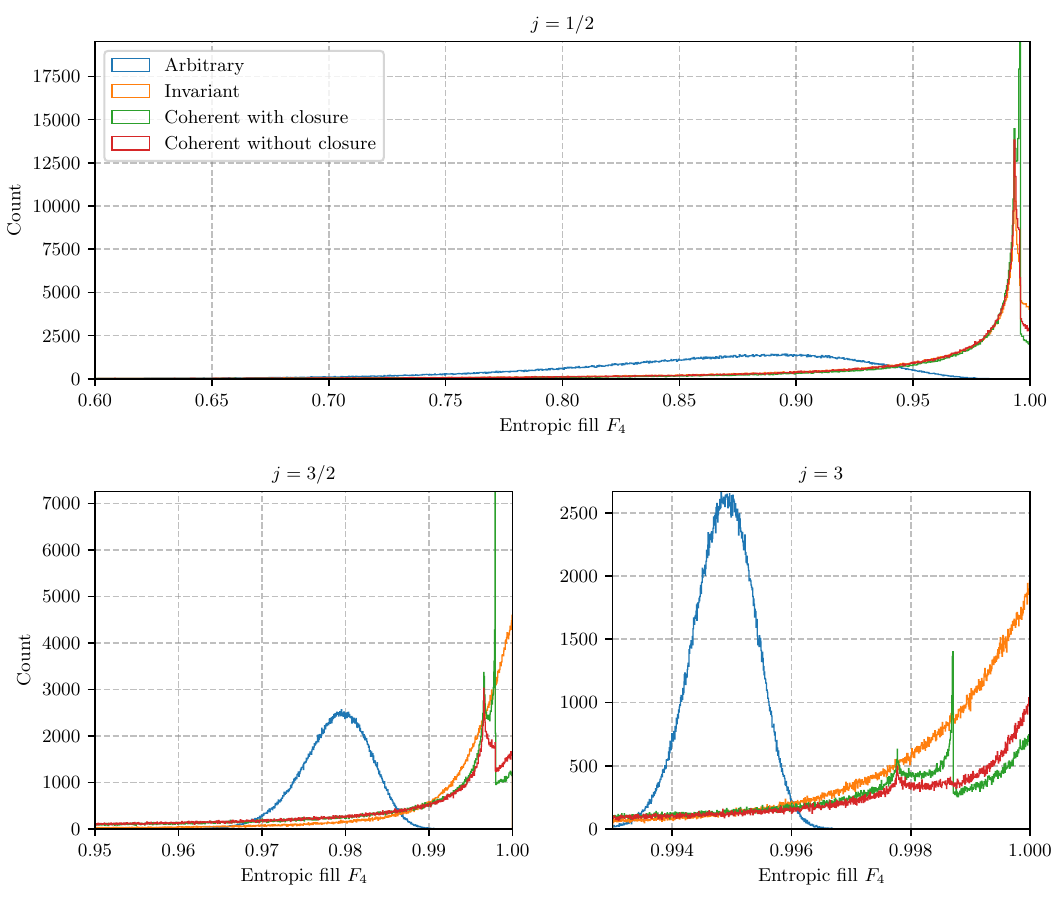}
    \caption{Sampled distributions of entropic fill for four-party arbitrary states, general intertwiners, and coherent intertwiners with and without closure (sample size \(5\times 10^5\) for each). We show \(1000\) bins in the interesting parts near maximal entropic fill (note the bounds on the \(x\)-axis for each plot) for \(j\in\{0.5,1.5,3\}\). Below these regions, the bins have comparably negligible counts. Average entropic fill of arbitrary states and general intertwiners increases more rapidly with increasing \(j\) than both types of coherent intertwiners, explaining the apparent change in relative counts from one plot to the next.}
    \label{fig:Distributions}
\end{figure*}

\begin{figure*}
    \centering
    \includegraphics[width=\textwidth]{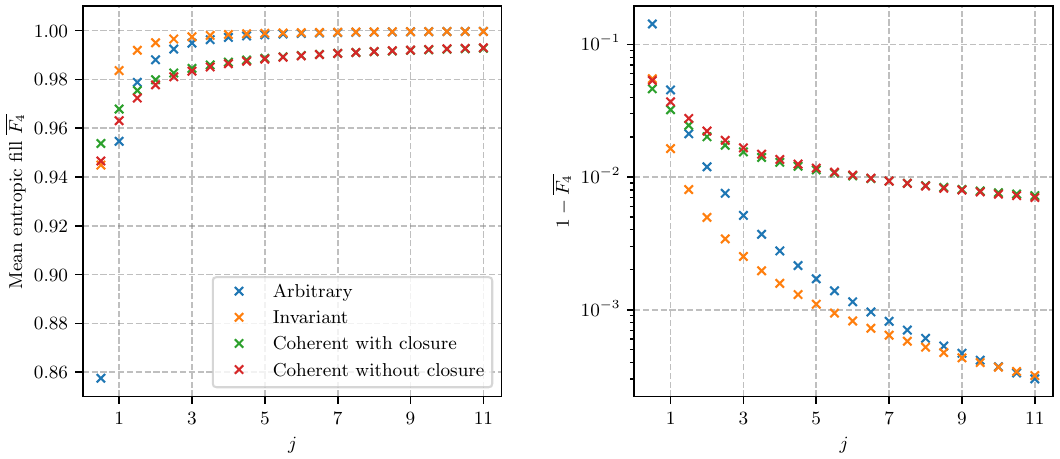}
    \caption{Numerically computed means of entropic fill for four-party arbitrary states, general intertwiners, and coherent intertwiners with and without closure (sample size \(5\times 10^5\) for each), for \(j\) ranging from \(1/2\) to \(11\). We show absolute entropic fill on the left, and the distance from maximal entropic fill on the right (logarithmic scale). All visible differences are well within statistical certainty, the standard errors of the means being at least an order of magnitude smaller at this sample size (except for the two coherent intertwiner categories between \(j=6.5\) and \(j=7.5\) and the two other categories near $j=10$).  The mean entanglement increases fastest for arbitrary states and general intertwiners, out-performing coherent intertwiners. Within the coherent category, the closure condition leads to slightly higher entropic fill for spin values up to about \(j=7\).}
    \label{fig:Distributions_means}
\end{figure*}

We compared entanglement distributions for several ensembles of four-valent tensors: arbitrary ones, generic invariant ones, and of coherent invariant tensors with and without closure for the classical data. While average entanglement is generally quite high and approaching its maximum for large $j$ in all ensembles as expected, the shape of the distributions is surprisingly different in the different ensembles (Fig.~\ref{fig:Distributions}). Mean entanglement grows at different rates in the different ensembles and seems to be highest in the ensemble of arbitrary tensors for large $j$ (Fig.~\ref{fig:Distributions_means}). Entanglement of invariant tensors peaks at the maximum possible value, whereas arbitrary tensors have a well-defined peak below maximum (Fig.~\ref{fig:Distributions}). The distribution for coherent intertwiners shows a lot of structure, for example two sharp peaks for the coherent intertwiners with closure, and one sharp peak for those without (Fig.~\ref{fig:Distributions}). It is interesting that these two ensembles seem to behave so differently, at least for low spin.

Note that the sampling for the ensembles of coherent intertwiners is uniform in their parameter space, but does not make reference to the Fubini-Study volume. This is a natural choice, but comparison with the other ensembles is then not necessarily straightforward. It could be interesting to also implement a different sampling, perhaps by pulling back the Fubini-Study metric to the space of coherent intertwiners.

\begin{figure*}
    \centering
    \includegraphics[width=\textwidth]{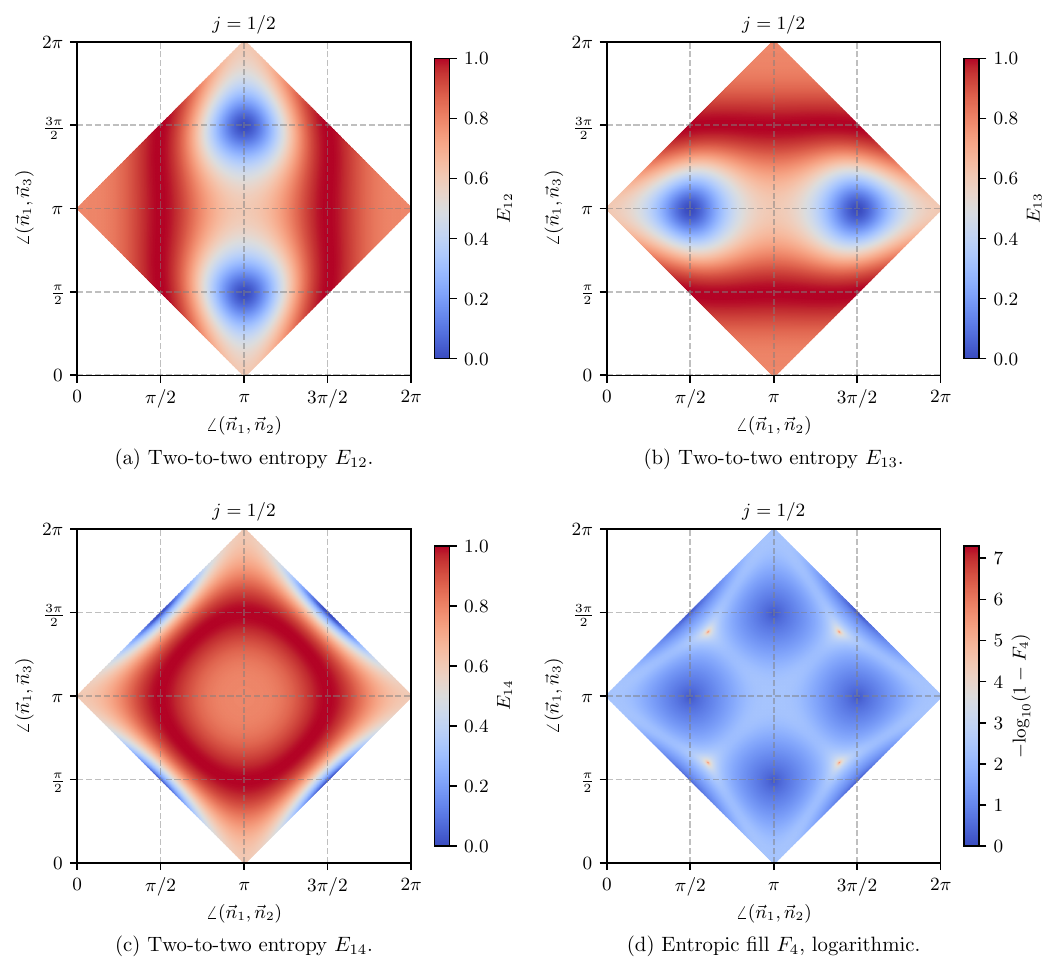}
    \caption{Numerically computed (\(2000\times 2000\)) two-to-two entropies and (logarithmic inverse distance from maximal) entropic fill on the configuration space of four-valent coherent intertwiners with closure for \(j=1/2\); the \(x\)-axis encodes the oriented angle~\(\theta_{12}\) between \(\vec{n}_1\) and \(\vec{n}_2\), while the \(y\)-axis represents the oriented angle~\(\theta_{13}=\angle(\vec{n}_1,\vec{n}_3)\). For closing configurations, these angles are restricted by \(\left|\theta_{12}-\pi\right|+\left|\theta_{13}-\pi\right|\le \pi\). We plotted both oriented angles ranging from zero to \(2\pi\) (at maximum), for clarity, which technically represents twice the configuration space of disphenoids. Indeed, we see more clearly in this parametrization that the measures of entanglement only depend on the cosines of the \(\angle(\vec{n}_i,\vec{n}_j)\). For the entropies, we used the analytical expressions~\eqref{eq:2-2entropiesspin0.5}, and entropic fill was determined numerically from these values. The regular tetrahedron configurations appear as spots of maximal entropic fill, but there are also degenerate points of vanishing \(F_4\), namely at (\(\theta_{12}=\pi,\ \theta_{13}\in\{\pi/2,3\pi/2\}\)) and index permutations thereof. Not every area element in these plots is equally likely for a random coherent intertwiner with closure.}
    \label{fig:CohClosure_FullConfigSpace_TH12_TH13_j0.5}
\end{figure*}

\begin{figure*}
    \centering
    \includegraphics[width=\textwidth]{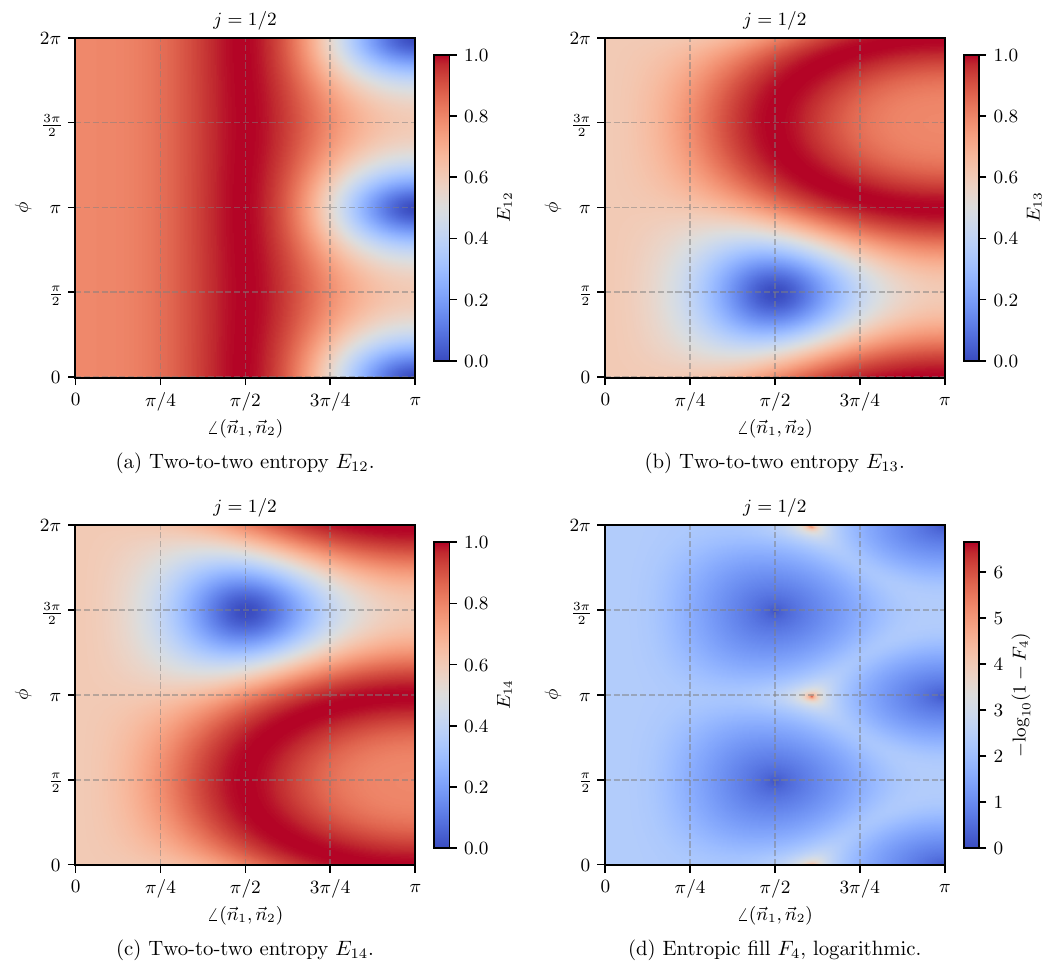}
    \caption{Numerically computed (\(300\times 300\)) two-to-two entropies and (logarithmic inverse distance from maximal) entropic fill on the configuration space of four-valent coherent intertwiners with closure for \(j=1/2\); the \(x\)-axis encodes the angle~\(\theta\) between \(\vec{n}_1\) and \(\vec{n}_2\), while the \(y\)-axis represents the angle~\(\phi\) on the circle of possible endpoints for \(\vec{n}_3\) given \(\vec{n}_1\) and~\({\vec{n}_2}\). The regular tetrahedron configurations appear as spots of maximal entropic fill. Not every area element in these plots is equally likely for a random coherent intertwiner with closure; in fact, \(\theta\) near \(\pi\) is more likely than near zero.}
    \label{fig:CohClosure_FullConfigSpace_j0.5}
\end{figure*}

\begin{figure*}
    \centering
    \includegraphics[width=\textwidth]{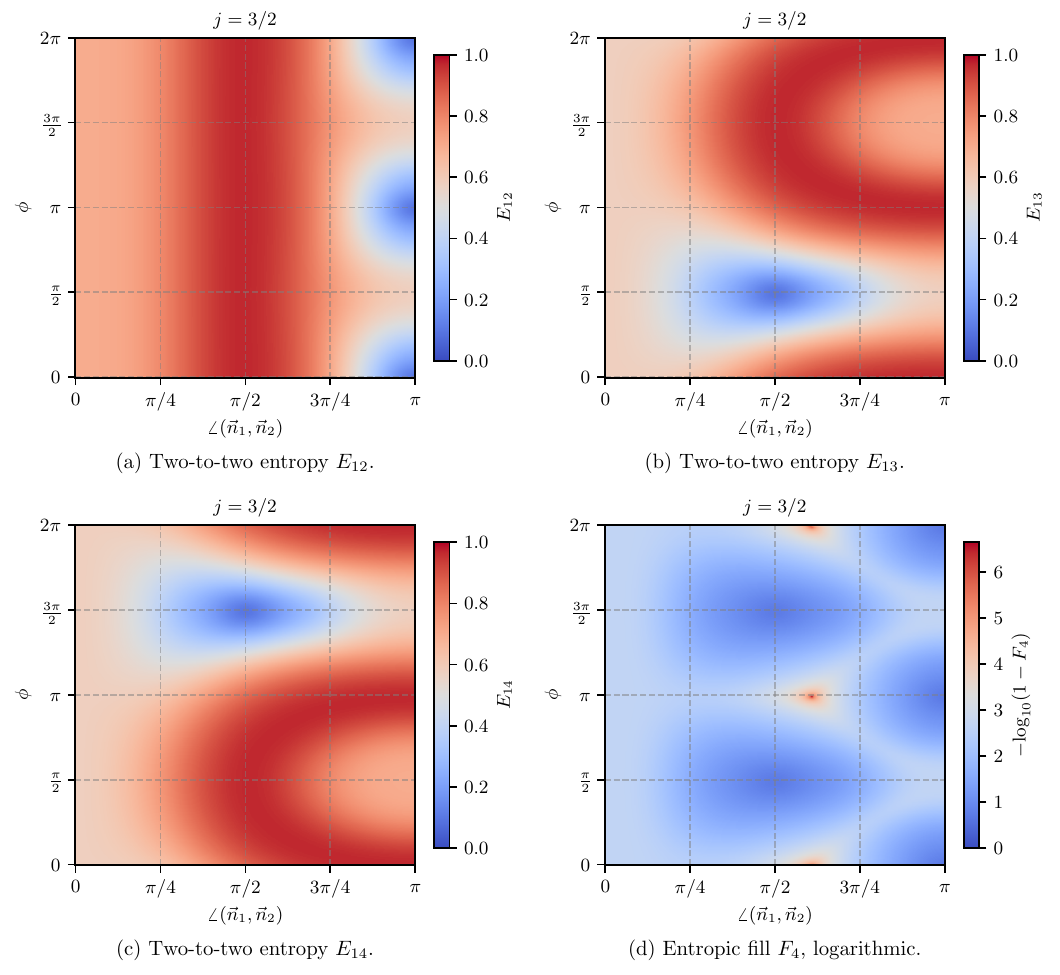}
    \caption{Idem Fig.~\ref{fig:CohClosure_FullConfigSpace_j0.5} with \(j=3/2\).}
    \label{fig:CohClosure_FullConfigSpace_j1.5}
\end{figure*}

\begin{figure*}
    \centering
    \includegraphics[width=\textwidth]{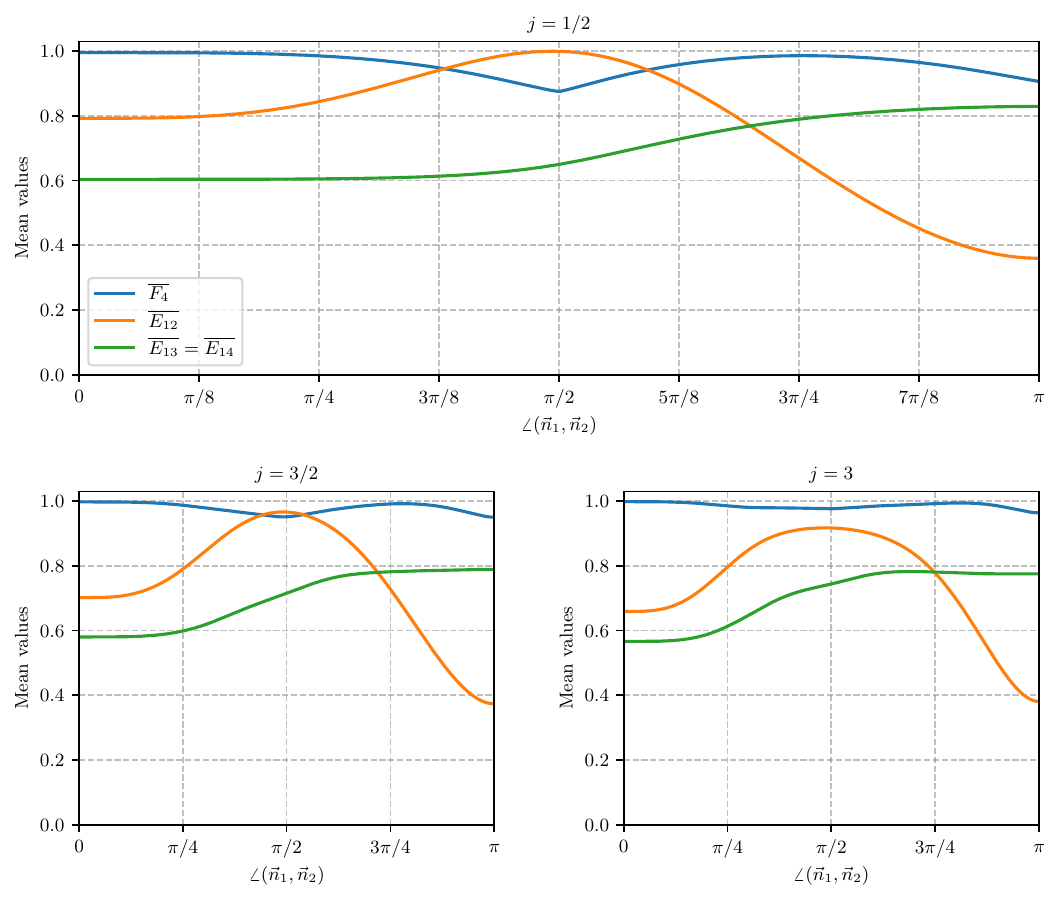}
    \caption{Numerically computed mean two-two-to entropies and entropic fill of four-valent coherent intertwiners with closure, as a function of the angle~\(\theta\) between \(\vec{n}_1\) and \(\vec{n}_2\), for \({j\in\{0.5,1.5,3\}}\). We sampled \(500\) equally spaced values of \(\phi\) for each of \(300\) equally spaced values of~\(\theta\).}
    \label{fig:Means_given_theta}
\end{figure*}

For the coherent intertwiners we also investigated the dependence of multipartite entanglement on the classical geometry that they approximate. The resulting picture is complex (Figs.~\ref{fig:CohClosure_FullConfigSpace_TH12_TH13_j0.5}--\ref{fig:Means_given_theta}): Multipartite entanglement depends on the geometry of the quantum tetrahedron in an intricate way. Generally speaking, entanglement is largest for the coherent intertwiner encoding the symmetric tetrahedron, and small for those which encode certain degenerate configurations.

For common spin \(j=1/2\), we expressed the two-to-two entropies of four-valent coherent intertwiners explicitly as functions of the angles enclosed by the unit normals parametrizing the state, and the method can be repeated for arbitrary \(j\) with increasingly tedious work. We found that for closed configurations, entropic fill vanishes only at particular, asymptotically degenerate points \eqref{eq:F4is0}. In other words, spin-\(1/2\) closed coherent intertwiners exhibit genuine multipartite entanglement for almost every configuration, and vanishing entanglement implies vanishing geometric volume of the tetrahedron parametrizing the coherent state. 

\begin{figure*}
    \centering
    \includegraphics[width=\textwidth]{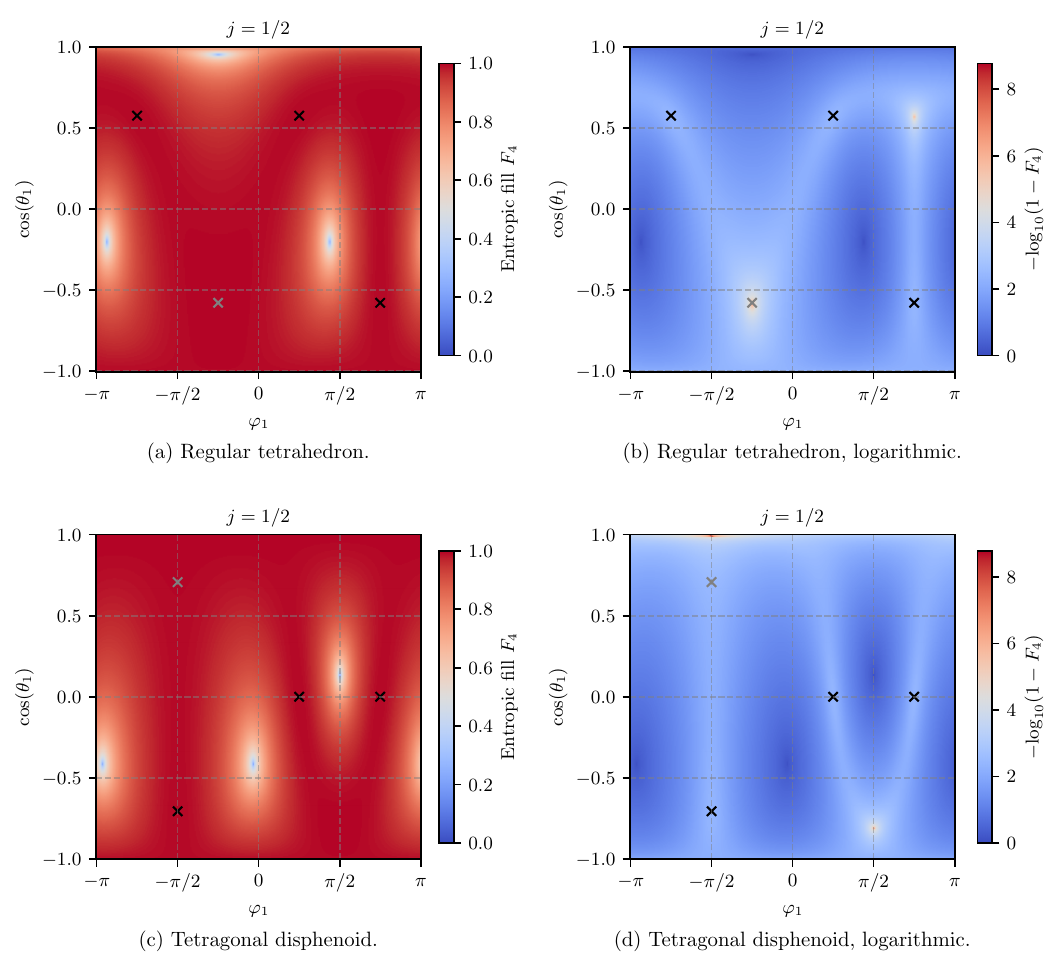}
    \caption{Numerically computed entropic fill for the regular tetrahedron and the tetragonal disphenoid base configurations for \(j=1/2\), as a function of the azimuthal angle \(\varphi_1\) and the cosine of the polar angle \(\cos(\theta_1)\) of \(\vec{n}_1\). Black crosses denote the coordinates of the other three unit normals; a gray cross is placed at the coordinates of the closure condition for \(\vec{n}_1\). We show absolute entropic fill on the left, and logarithmic inverse distance from maximal entropic fill on the right. In these plots, every area element is equally likely.}
    \label{fig:RegTet_Disph_j0.5}
\end{figure*}

\begin{figure*}
    \centering
    \includegraphics[width=\textwidth]{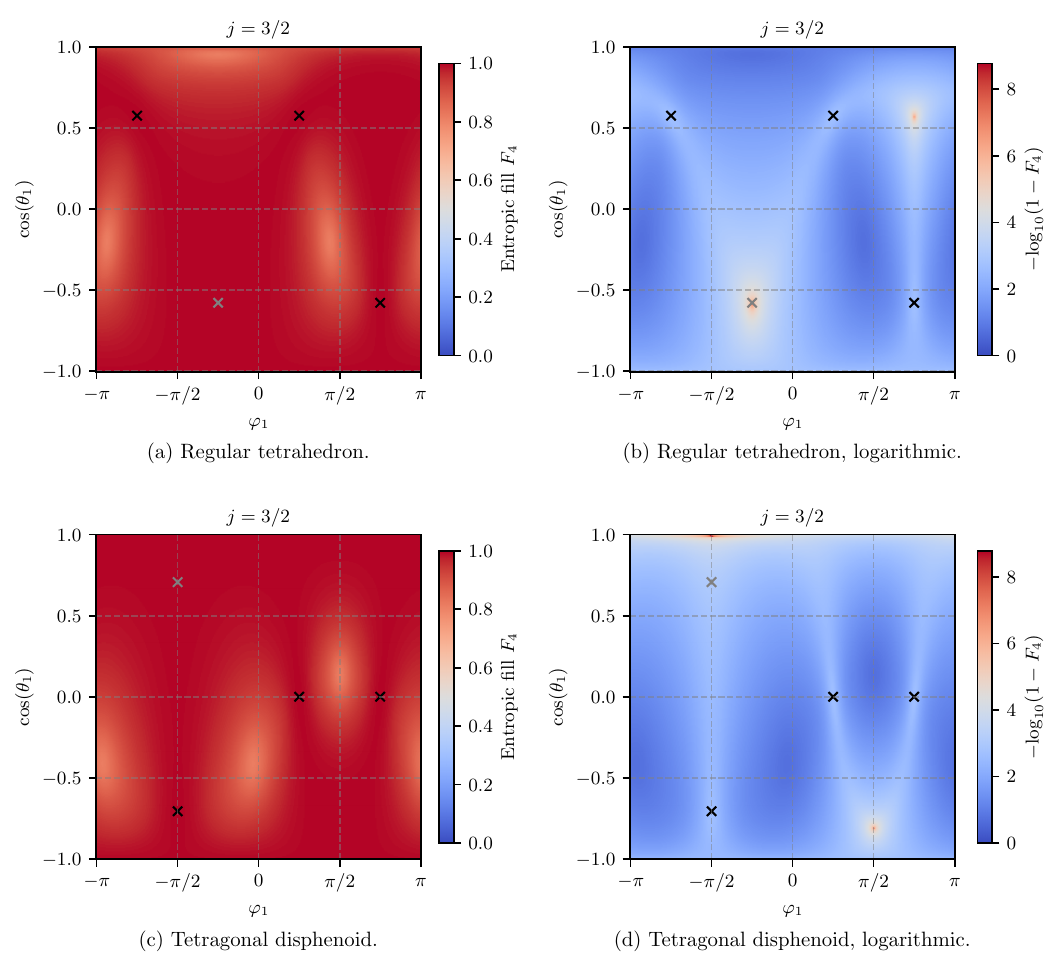}
    \caption{Idem Fig.~\ref{fig:RegTet_Disph_j0.5} with \(j=3/2\).}
    \label{fig:RegTet_Disph_j1.5}
\end{figure*}

Regarding the effects of (non-)closure of the classical data, the situation is similarly intriguing. The entanglement is higher in the closed configuration than in a typical non-closed one, but the closed configuration is not always the location of maximal entanglement (Figs.~\ref{fig:RegTet_Disph_j0.5},~\ref{fig:RegTet_Disph_j1.5}).

We should note that, by solving Eqs.\ \eqref{eq:F4FirstSet}, \eqref{eq:F4SecondSet} for a large number of intertwiners and thus associating an entropic tetrahedron in each case, we have added to the strong numerical evidence \cite{EntropicFill} that entropic fill is defined for any local dimension. 

As a final point of discussion, both the quantum geometry and the entanglement measure $F_4$ studied in this work are related to tetrahedra. A natural question is whether, and how, the geometries of the entropic tetrahedron determined by $\eqref{eq:F4FirstSet}, \eqref{eq:F4SecondSet}$ and the geometric tetrahedron parametrizing the coherent state are related. 

The face areas of the entropic tetrahedron are determined by the (normalized) one-to-other von Neumann entropies. For intertwiners it is maximal, $\log_2 d_j$ without normalization for a spin $j$ party. The area of the geometric triangle also depends on $j$, albeit in a different way, $a_j \propto \sqrt{j(j+1)}$. Note, however, that if one of the tetrahedra degenerates because a face area becomes zero, then so does the other. 

For the rest of the degrees of freedom of a tetrahedron, the relationship is more complex and indirect, see Figs.~\ref{fig:CohClosure_FullConfigSpace_j0.5} and~\ref{fig:CohClosure_FullConfigSpace_j1.5}. Extrapolating from the data presented in this work, one can say that maximum volume of the entropic tetrahedra is assumed for a symmetric geometric tetrahedron, and minimum volume of the entropic tetrahedra seems to imply a degenerate geometric tetrahedron. Finally, violating geometric closure typically lowers the volume of the entropic tetrahedron, at least for low \(j\).
It would be very interesting to use the powerful spinorial description \cite{Freidel:2010xna,Freidel:2010aq,Freidel:2010tt,Borja:2010rc} 
of coherent intertwiners to obtain more general results of this kind.

In loop quantum gravity, quantum states have geometrical meaning \cite{Thiemann:2001gmi,Ashtekar:2004eh}. This ties into a larger story about entanglement in quantum gravity \cite{Srednicki:1993im,Bombelli:1986rw,Perez:2014ura,Bianchi:2012ev}. The present work sheds light on just a small aspect of the correlation between geometric properties and properties of the states of loop quantum gravity in terms of quantum information: Our work shows how the geometry of a single atom of space, the quantum tetrahedron, and its entanglement properties as a part of the quantum state of spacetime geometry, go hand in hand: Maximal and minimal entropy are attained for geometrically special states, geometric closure influences entanglement, and ensembles of the semiclassical coherent intertwiners  behave very differently from that of general intertwiners, at least as far as we were able to probe.

 As a direct continuation of the present work, it would be interesting to better understand the structures seen in the entanglement of coherent intertwiners. More generally, long-range entanglement in loop quantum gravity results from states that are not product states with respect to the intertwiners at different vertices~\cite{Livine:2017fgq,Baytas:2018wjd,Bianchi:2018fmq}. It would be very interesting to understand this type of entanglement and its relation to the geometry better, in particular also for solutions to the quantum constraints, a form of Einstein's equations in the quantum theory~\cite{Thiemann:2001gmi,Ashtekar:2004eh}.

\section*{Acknowledgments}

We thank the anonymous referees for their questions and comments. They helped improve the manuscript. 
HS acknowledges the contribution from COST Action BridgeQG (CA23130), supported by COST (European Cooperation in Science
and Technology).

\section*{Data and code availability}

All code used and data created for this publication is available on GitHub~\cite{IntertwinerEntFillGitHub}.

\bibliography{references}

\end{document}